\documentclass[twocolumn, superscriptaddress]{revtex4}

\usepackage{amssymb}
\usepackage{amsmath}
\usepackage{graphicx}
\usepackage{latexsym}

\def\Slash#1{{\ooalign{\hfil$#1$\hfil\crcr\hfil$/$\hfil}}}

\begin{document}

\title{Fate of the False Vacuum Revisited}

\author{Shigeki Matsumoto}
\affiliation{Department of Physics, University of Toyama, Toyama 930-8555, Japan}
\author{Keiko I. Nagao}
\affiliation{Department of Physics, National Tsing Hua University, Taiwan 30013, R.O.C.}
\affiliation{Department of Physics, Nagoya University, Nagoya 464-8602, Japan}
\author{Makoto Nakamura}
\affiliation{Department of Physics, University of Toyama, Toyama 930-8555, Japan}
\author{Masato Senami\footnote{This author is now in Department of Micro Engineering, Kyoto University, Kyoto 606-8501, Japan}}
\affiliation{ICRR, University of Tokyo, Kashiwa, Chiba 277-8582, Japan}

\begin{abstract}
We find the novel effect on the decay of a false vacuum
in view of quantum field theory,
which is induced by a field coupling to the scalar field 
related to a first-order phase transition.
This effect of the environment can never be included in
the traditional method using the effective potential,
and, in fact, acts as dissipative and fluctuation effects on tunneling phenomena.
We show that the decay of the false vacuum is drastically either enhanced or suppressed.
It is also clarified what kind of interaction enhance or suppress the tunneling probability. 
\end{abstract}



\maketitle 

\section{Introduction}

For the last three decades,  
the first-order phase transition remains one of the most important topics
in the cosmology, particle physics, and condensed matter.
In particular, this topic is widely related to many phenomena about the early universe,
e.g., the electroweak phase transition \cite{EWbaryogenesis},
the quark confinement~\cite{QGP},
and inflationary models \cite{inflation}.
Since the phase transition is the phenomena closely related to bosonic degree,
it is an inevitably important element for supersymmetric models.
Recently, this topic is discussed
in relation to charge and/or color breaking minima
and supersymmetry breaking metastable minima,
which is intensively studied after the recent work by Intriligator et al.~\cite{Intriligator:2006dd}.

For first-order phase transitions, 
the decay rate of a false vacuum is the key ingredient.
The estimate of the rate has been studied in many works,
which is usually calculated using the Euclidean action of the bounce solution~\cite{Coleman:1977py,Coleman:1977th}.
The accuracy of the estimate has also been improved
with considering the effects of dissipation and finite temperature~\cite{dissipation,Linde:1981zj}.

In this Letter, we find that a novel correction to the effective action
in view of quantum field theory,
which is induced by a field coupling to the scalar field
related to a first-order phase transition.
Hence, this effect is considered to be induced by the environment.
This correction cannot be treated in the method of the effective potential
and acts as dissipative and fluctuation effects on tunneling phenomena.
For a phase transition induced by classical thermal fluctuation,
some effects of environment have been studied so far.
For the quantum tunneling,
the effect of the environment has been studied in viewpoints of quantum mechanics.
We study this effect in a framework of quantum field theory
and find a novel correction.
This effect changes drastically the decay rate of the false vacuum,
since it is the correction to the action,
where the decay rate depends on the action exponentially.
It is also shown that this effect drastically makes the lifetime of the false vacuum 
either longer or shorter depending on the interaction which couples the scalar field.

\section{Tunneling without Environment}

\begin{figure}[t]
\scalebox{0.7}{\includegraphics{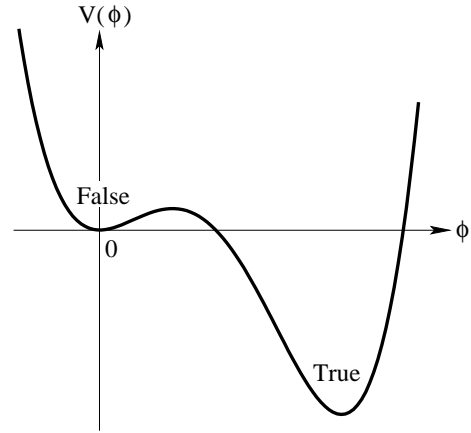}}
\caption{\small Potential describing the tunneling of the $\phi$ field.}
\label{fig: Potential}
\end{figure}

We briefly summarize how we calculate the probability of the tunneling without the environment using the bounce method~\cite{Coleman:1977py} and clarify our notation. With the potential shown in Fig.~\ref{fig: Potential}, the Lagrangian which describes a first-order phase transition of the $\phi$ field is given by
\begin{eqnarray}
{\cal L}_\phi
=
\frac{1}{2}
\left(\partial_\mu \phi \right)
\left(\partial^\mu \phi \right)
-
V(\phi).
\label{Lagrangian without environment}
\end{eqnarray}
The tunneling probability from the false vacuum to the true one, per unit time per unit volume, in the semi-classical approximation is given by the following formula,
\begin{eqnarray}
\Gamma/V \simeq K \exp\left(-{\cal S}_{\rm E}[\phi_{\rm B}]\right),
\end{eqnarray}
where the configuration $\phi_{\rm B}$ is called the bounce solution, and it satisfies the equation of motion of the Euclidean action ${\cal S}_{\rm E}[\phi]$ deduced from the Lagrangian (\ref{Lagrangian without environment}),
\begin{eqnarray}
-\left(\frac{d^2}{dr^2} + \frac{3}{r}\frac{d}{dr}\right)\phi_{\rm B}
+ V^\prime (\phi_{\rm B}) = 0.
\end{eqnarray}
Here, the prime denotes the derivative by the argument. The O(4) symmetry of the solution, $\phi_{\rm B}(x_{\rm E}) = \phi_{\rm B}(r = |x_{\rm E}|)$, is assumed in the above equation, where $x_{\rm E}$ is the coordinates of the Euclidean space-time~\cite{Coleman:1977th}. The solution also satisfies the boundary conditions, $\phi_{\rm B}(\infty) = 0$ and $\dot{\phi}_{\rm B}(0) = 0$, where the dot indicates the derivative by $r$.

On the other hand, the coefficient $K$ has the form of
\begin{eqnarray}
K = \frac{{\cal S}_{\rm E}[\phi_{\rm B}]^2}{4\pi^2}
\sqrt{
\left|
\frac{\det[-\partial_\mu\partial_\mu + V^{\prime\prime}(0)]}
{\det^\prime[-\partial_\mu\partial_\mu + V^{\prime\prime}(\phi_{\rm B})]}
\right|},
\label{Coefficient K}
\end{eqnarray}
where the primed determinant means that the zero modes of the operator, $-\partial_\mu\partial_\mu + V^{\prime\prime}$, which correspond to the translational invariance about the bounce's location, is subtracted. Instead of the subtraction, the bounce action ${\cal S}_{\rm E}[\phi_{\rm B}]$ appears in the right-hand side of Eq.~(\ref{Coefficient K}).

\section{Tunneling with Environment}

We discuss how the effect of the environment, namely, interactions between $\phi$ and other fields alter the tunneling probability. For quantitative discussion, we consider two specific cases described by the following Lagrangians,
\begin{eqnarray}
{\cal L}_{\rm tot}
&=&
{\cal L}_\phi
+
{\cal L}^{(I)}_{\rm env}
+
{\cal L}_c,
\label{Lagrangian with environment}
\\
{\cal L}_{\rm env}^{(S)}
&=&
-S^*(\Box + m_S^2 + y_S M \phi)S,
\label{Lagrangian S}
\\
{\cal L}_{\rm env}^{(F)}
&=&
+\bar{F}\left(i\Slash{\partial} - m_F - y_F \phi \right) F,
\label{Lagrangian F}
\end{eqnarray}
where $(I)$ is $(S)$ or $(F)$, and the field $S$ ($F$) is a complex scalar boson (a Dirac fermion) with $m_S$ ($m_F$) being its mass, and $M$ is the curvature of the potential minimum at $\phi = 0$, namely, the mass of $\phi$ in the false vacuum. The strength of the interaction between $\phi$ and $S$ ($F$) is characterized by the coupling constant $y_S$ ($y_F$). The counter terms ${\cal L}_c$ are introduced in the Lagrangians, which regularize the environmental effect of the interactions. The on-shell renormalization condition is adopted for ${\cal L}_c$.

The probability of the tunneling with the environment is obtained from the effective Euclidean action. This action is derived by integrating the environment field ($S$ or $F$) out from the total Euclidean action deduced from the Lagrangian (\ref{Lagrangian with environment}). This action has the form of
\begin{align}
{\cal S}_{\rm E}^{(I)}[\phi]
=&
\int d^4x_{\rm E} \,
\left[
\frac{1}{2} \dot{\phi}^2
+
\frac{1}{2} \left(\nabla \phi\right)^2
+
V(\phi)
\right]
\label{Effective Euclidean action}
\\
&
+\frac{1}{2}
\int d^4x_{\rm E} \int d^4x^\prime_{\rm E} \,
\phi(x_{\rm E})
f_0^{(I)}(x_{\rm E} - x^\prime_{\rm E})
\phi(x^\prime_{\rm E}),
\nonumber
\end{align}
where we implicitly assume that coupling constants ($y_S$ and $y_F$) are small enough and hence neglect higher order terms of $\phi$ than $\phi^2$. The effect of the environment is summarized in the function $f_0^{(I)}(x_{\rm E})$, and, for later use, it is convenient to express it in the Fourier transform,
\begin{eqnarray}
f_0^{(I)}(x_{\rm E})
=
\int \frac{d^4p}{(2\pi)^4} \tilde{f}^{(I)}(p) e^{-i p x_{\rm E}}
+
\tilde{f}_0^{(I)}(0) \, \delta^{(4)}(x_{\rm E}),
\nonumber
\end{eqnarray}
where $\tilde{f}^{(I)}(p) = \tilde{f}_0^{(I)}(p) - \tilde{f}_0^{(I)}(0)$. The term proportional to a delta-function gives a finite correction to the potential $V(\phi)$, which is renormalized by the redefinition of the mass term in the potential. On the other hand, the term involving $\tilde{f}^{(I)}(p)$ cannot be a correction to $V(\phi)$, because this term vanishes when $\phi$ is independent of $x_{\rm E}$. In fact, when we expand $\phi(x_{\rm E}) \int d^4x^\prime_{\rm E} f^{(I)}(x_{\rm E} - x^\prime_{\rm E}) \phi(x^\prime_{\rm E})$ in terms of local operators including two $\phi$'s, these operators have to involve the derivative of $\phi$ such as $\phi \Box \phi$. This effect of the environment, therefore, is never taken into account in the method of the effective potential. In this Letter, we focus mainly on how this effect alters the tunneling probability.

Explicit expression of the function $\tilde{f}^{(I)}(p)$ obtained from the interactions (\ref{Lagrangian S}) and (\ref{Lagrangian F}) are given by
\begin{eqnarray}
\tilde{f}^{(I)}(p)
&=&
\int^\Lambda_{2m_I} d\omega
\frac{p^2 \, r^{(I)}(\omega)}{\omega^2(\omega^2 + p^2)}
-
p^2 \Pi^{(I)\prime}(M^2),
\label{Function f}
\\
r^{(S)}(\omega)
&=&
\frac{\alpha_S}{2\pi} M^2 \left(\omega^2 - 4m_S^2\right)^{1/2},
\label{r of S}
\\
r^{(F)}(\omega)
&=&
\frac{\alpha_F}{\pi} \left(\omega^2 - 4m_F^2\right)^{3/2},
\label{r of F}
\end{eqnarray}
where $\alpha_I = y_I^2/(4\pi)$ and $\Lambda$ is the cutoff parameter, which is eventually taken to be infinity. The function $\Pi^{(I)}(M^2)$ comes from the counter terms ${\cal L}_c$, which is defined by
\begin{eqnarray}
\Pi^{(I)}(M^2)
=
{\cal P} \int^\Lambda_{2m_I} d\omega \, \frac{r^{(I)}(\omega)}{\omega^2 - M^2},
\end{eqnarray}
where ${\cal P}$ denotes a principal valued integral.

The equation of motion for the bounce solution with the environment is obtained from the effective Euclidean action (\ref{Effective Euclidean action}). Imposing the O(4) symmetry on the solution, the equation turns out to be
\begin{eqnarray}
&&
-\left(\frac{d^2}{dr^2} + \frac{3}{r}\frac{d}{dr}\right)\phi^{(I)}_{\rm B}
+
V^\prime (\phi^{(I)}_{\rm B})
\label{EOM with environment}
\\
&&
+ \int dr' \int dp \,
\frac{p \, r^{\prime 2} J_1(pr) J_1(pr')}{r}
\tilde{f}^{(I)}(p) \phi^{(I)}_{\rm B}(r') = 0,
\nonumber
\end{eqnarray}
with the boundary conditions, $\phi^{(I)}_{\rm B}(\infty) = 0$ and $\dot{\phi}^{(I)}_{\rm B}(0) = 0$. Here, $J_1(x)$ is the Bessel function of the first kind. Once the bounce solution $\phi^{(I)}_{\rm B}$ is computed by solving the above integro-differential equation, the probability of the tunneling with the environment is obtained as
\begin{eqnarray}
\Gamma(\alpha_I)/V
\simeq
K^{(I)} \exp \left(-{\cal S}^{(I)}_{\rm E}[\phi^{(I)}_{\rm B}]\right).
\label{Probability with environment}
\end{eqnarray}
Since we are interested in cases where interactions between $\phi$ and environment fields are weak enough, the effect of the environment in the coefficient $K^{(I)}$ is weak, at most, ${\cal O}$(10)\% level. We therefore neglect such a correction in the following discussion. On the other hand, the effect in ${\cal S}^{(I)}_{\rm E}[\phi^{(I)}_{\rm B}]$ can be very significant because of its exponential sensitivity as seen in the probability (\ref{Probability with environment}).

\begin{figure}[t]
\hspace{-8mm}
\scalebox{0.7}{\includegraphics{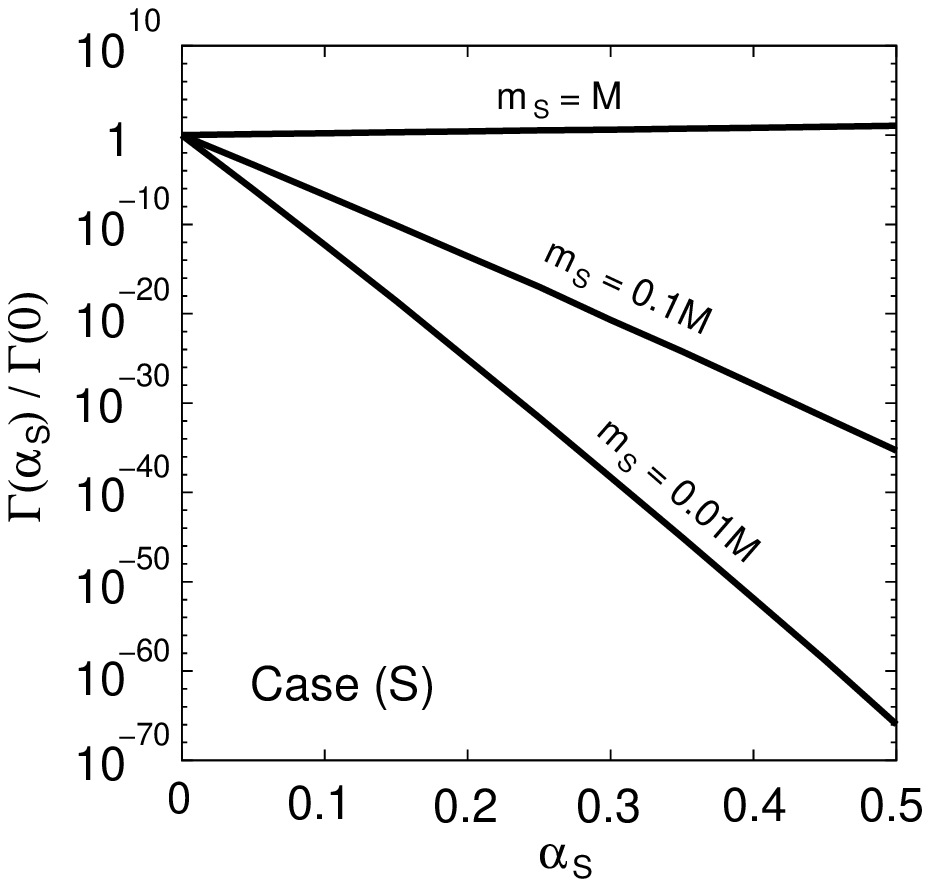}}
\scalebox{0.7}{\includegraphics{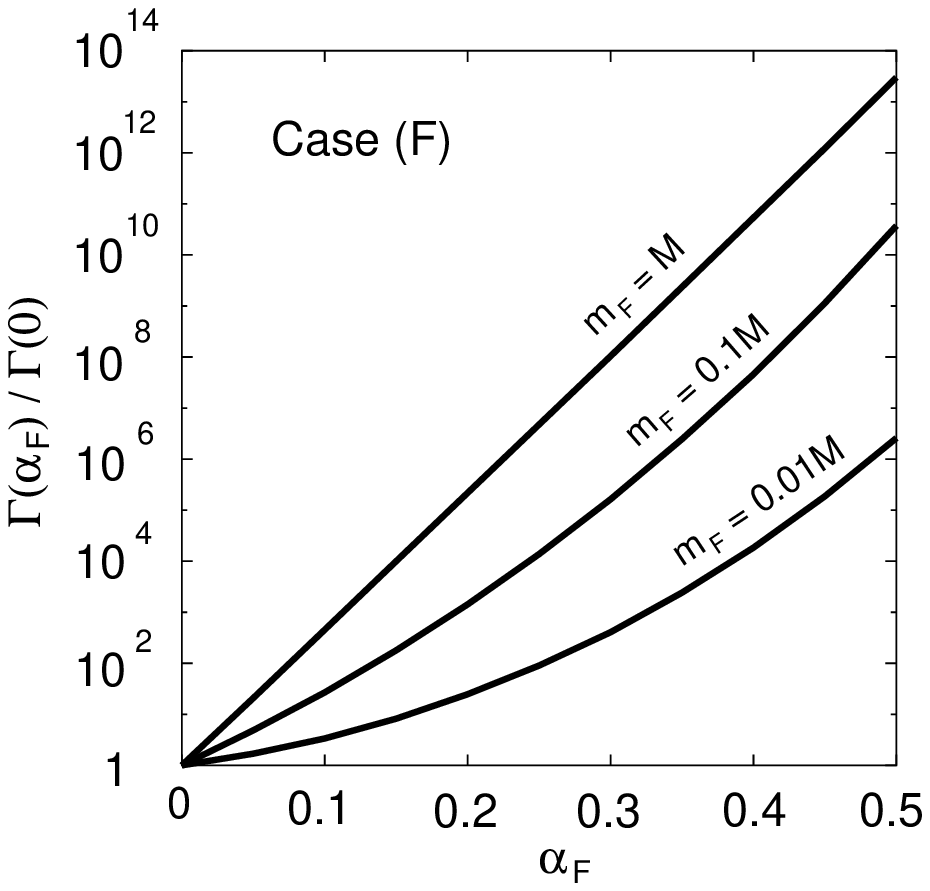}}
\qquad\qquad
\caption{\small Ratio of the tunneling probability between with and without the environment.}
\label{fig: Tunneling ratio}
\end{figure}

In Fig.~\ref{fig: Tunneling ratio}, the ratio of the tunneling probability between with and without the environment is shown as a function of $\alpha_I$ with several choices of $m_I$. A cubic potential,
\begin{eqnarray}
V(\phi)
=
\frac{1}{2} M^2 \phi^2 \left(1 - \frac{\phi}{\phi_0}\right),
\label{Cubic potential}
\end{eqnarray}
is used to calculate the ratio. The metastable well and the potential barrier are assumed to be well approximated by the above potential. The parameter $\phi_0/M$ is fixed so that the bounce action without the environment takes the value of 400, which is required for the stability of the false vacuum. As can be seen in the figure, the effect of the environment becomes significant when $m_I$ is small, especially, in the case $(S)$.

\section{Effect of Environment}

We are now in position to discuss the effect of the environment shown in Fig.~\ref{fig: Tunneling ratio} in detail. For the intuitive understanding of the effect, it is convenient to rewrite the integro-differential equation (\ref{EOM with environment}) to a simple integral equation with the help of the following transformations,
\begin{eqnarray}
\phi^{(I)}_{\rm B}(r)
&=&
\int^\infty_0 dp \, \frac{p}{r} J_1(pr) \, \tilde{\phi}^{(I)}_{\rm B}(p),
\\
\tilde{\phi}^{(I)}_{\rm B}(p)
&=&
\int^\infty_0 dr \, r^2 J_1(pr) \, \phi^{(I)}_{\rm B}(r).
\end{eqnarray}
It is easy to confirm that the above transformations satisfy the boundary conditions for $\phi^{(I)}_{\rm B}$. It then turns out that Eq.~(\ref{EOM with environment}) is translated into
\begin{eqnarray}
&&
\frac{\pi \phi_0}{3M^2}
\left[p^2 + M^2 + \tilde{f}^{(I)}(p)\right]
\tilde{\phi}^{(I)}_{\rm B}(p)
\label{Integral equation}
\\
&&
=
\int^\infty_0 dp' \int^{p + p'}_{|p - p'|} dp'' \,
\tilde{\phi}^{(I)}_{\rm B}(p') \, \tilde{\phi}^{(I)}_{\rm B}(p'')
\frac{\Delta(p, p', p'')}{p},
\nonumber
\end{eqnarray}
where the function $\Delta(p, p', p'')$ gives the area of the triangle with the sides $p$, $p'$, and $p''$. The cubic potential (\ref{Cubic potential}) is assumed in the above equation, while it is also possible to apply the translation to more general cases. Using the function $\tilde{\phi}^{(I)}_{\rm B}(p)$, the bounce action is given by
\begin{eqnarray}
{\cal S}_{\rm E}^{(I)}[\phi^{(I)}_{\rm B}]
&=&
\frac{\pi M^2}{\phi_0} \int dp \, dp' \, dp''
\label{Bounce action}
\\
&\times&
\tilde{\phi}^{(I)}_{\rm B}(p) \,
\tilde{\phi}^{(I)}_{\rm B}(p') \,
\tilde{\phi}^{(I)}_{\rm B}(p'') \,
\Delta(p, p', p'').
\nonumber
\end{eqnarray}

We first notice that $\tilde{\phi}^{(I)}_{\rm B}(p)$ is rapidly decreased when $p \gg M$, because Eq.~(\ref{Integral equation}) gives the relation $(\phi_0/M^2) \, p^2 \, \tilde{\phi}^{(I)}_{\rm B} \sim p^3 \, (\tilde{\phi}^{(I)}_{\rm B})^2$, which is obtained from a simple dimensional analysis. The function $\tilde{\phi}^{(I)}_{\rm B}(p)$ has, therefore, a non-negligible value in the region $p \lesssim M$. On the other hand, Eq.~(\ref{Integral equation}) also indicates, when the effect of the environment, $\tilde{f}^{(I)}(p)$, is larger, $\tilde{\phi}^{(I)}_{\rm B}(p)$ is larger in order to keep a balance between the right- and left-hand sides of Eq.~(\ref{Integral equation}). It leads to larger action (\ref{Bounce action}), namely, smaller tunneling probability.

\begin{figure}[t]
\hspace{-8mm}
\scalebox{0.7}{\includegraphics{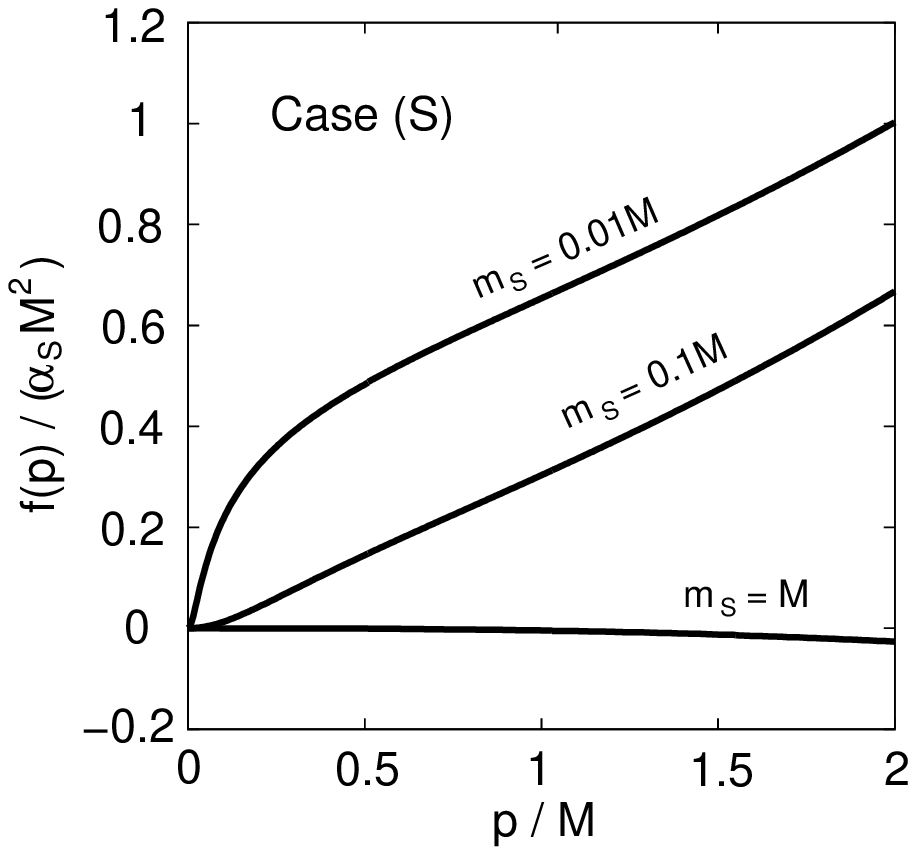}}
\scalebox{0.7}{\includegraphics{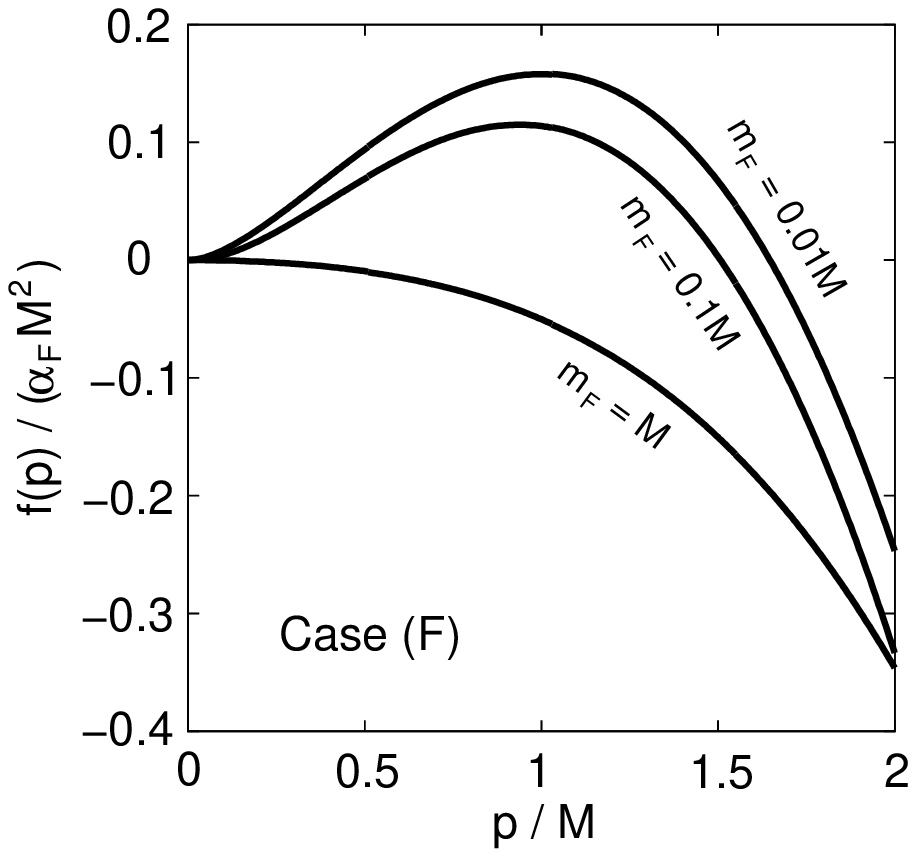}}
\qquad\qquad
\caption{\small Effect of the environment, $\tilde{f}^{(I)}(p)$,
in the unit of $\alpha_I M^2$.}
\label{fig: Environment effect}
\end{figure}

In Fig.~\ref{fig: Environment effect}, the effect of the environment, $\tilde{f}^{(I)}(p)$, is shown as a function of $p$ with several choices of $m_I$. It can be seen that the effect is larger when $m_I$ is smaller, while no enhancement can be found in the case of $m_I = M$. This fact can be intuitively understood, because the $\phi$ field cannot decay into environment fields when $M \leq 2m_I$ and the dissipative effect on the tunneling disappears. When $M > 2m_I$, this effect is more significant for smaller $m_I$ as expected. The difference between the cases $(S)$ and $(F)$ on this effect comes from the fact that the $\phi$ field decays to $SS^*$ through the S-wave process, while it decays to $F\bar{F}$ through the P-wave one. The function $\tilde{f}^{(I)}(p)$, therefore, grows quickly at $p \lesssim M$ in the case $(S)$, which leads to a strong dissipative effect on the tunneling as shown in Fig.~\ref{fig: Tunneling ratio}. On the other hand, in the case $(F)$, $\tilde{f}^{(I)}(p)$ grows slowly, which result in a small dissipative effect as also shown in Fig.~\ref{fig: Tunneling ratio}.

There is another important effect of the environment, which can be especially seen in the case $(F)$. The function $\tilde{f}^{(I)}(p)$ can be negative, leading to the suppression of the bounce action, namely, the enhancement of the tunneling probability. Since the effect does not vanish even in the case of $M = m_I$, it cannot be a dissipative effect. Instead, it should be regarded as a effect of fluctuation due to the interaction between $\phi$ and an environment field. This fact can be understood by considering the case $(F)$ at $m_F = 0$ limit, where $\tilde{f}^{(I)}(p)$ is analytically given by $\tilde{f}^{(F)}(p) = (\alpha_F/2\pi) [p^2 - p^2 \ln(p^2/M^2)]$. The logarithmic term is the origin of the suppression. The combination of $p^2 + \tilde{f}^{(F)}(p)$ in Eq.~(\ref{Integral equation}) is rewritten as $\sim M^2(p^2/M^2)^{1 - \alpha_F/(2\pi)}$, which is nothing but the anomalous dimension from the Yukawa interaction $y_F\phi\bar{F}F$. Since the existence of this effect and the weak dissipative effect in the case $(F)$, the tunneling probability is enhanced compared to that without the environment as shown in Fig.~\ref{fig: Tunneling ratio}.

\section{Summary}

We have investigated the effect of the environment on the tunneling probability which cannot be taken into account in the method of the effective potential. We have formulated how the effect is involved in the calculation of the tunneling probability within the framework of the bounce method and shown that the effect indeed can be very significant. We have also clarified what kind of interaction enhance or suppress the tunneling probability. It is, therefore, important to consider this effect for the estimation of the probability, especially in models built on a false vacuum. It may also be interesting to consider the scenario on a false vacuum which is stabilized using the environment discussed in this Letter.

\section*{Acknowledgements}

The work of S. M. was supported in part by the Grant-in-Aid for Scientific Research from the Ministry of Education, Culture, Sports, Science, and Technology (MEXT) (Nos.\ 21740174 and 22244021).
The work of K. I. N. was supported in part
by Grant-in-Aid for Nagoya University Global COE Program from the MEXT
and the Grant-in-Aid for JSPS Fellows.


\end{document}